\documentclass[traditabstract]{aa}
\usepackage{graphicx}
\usepackage{natbib}
\usepackage{txfonts}
\bibpunct{(}{)}{;}{a}{}{,}
\begin{document}

  \title{Lithium and oxygen in globular cluster dwarfs and the early disc accretion scenario}
  \author{Maurizio Salaris\inst{1} \and   
          Santi Cassisi\inst{2}}

\institute{Astrophysics Research Institute, Liverpool John Moores University, 
           IC2, Liverpool Science Park, 
           146 Brownlow Hill, 
           Liverpool L3 5RF, UK  \email{M.Salaris@ljmu.ac.uk}
           \and           
           INAF~$-$~Osservatorio Astronomico di Collurania, Via M. Maggini, I$-$64100 , Teramo, Italy 
            \email{cassisi@oa-teramo.inaf.it} 
           }
  \abstract{A new scenario --early disc accretion-- has been recently proposed to explain the discovery 
of multiple stellar populations in Galactic globular clusters. According to this model, 
the existence of well defined (anti)-correlations amongst light element abundances (i.e. C, N, O, Na) in the photospheres of stars belonging 
to the same cluster (and the associated helium enrichment), is caused by accretion of the ejecta of short lived 
interacting massive binary systems (and single fast rotating massive stars) 
on fully convective pre-main sequence low- and very low-mass stars, during the early stages of the cluster evolution.
In a previous paper we have applied this scenario to the cluster NGC2808, showing how the knowledge of the He abundance of its 
triple main sequence can constrain both the He abundance of the accreted matter 
and the accretion efficiency. 
Here we investigated the constraints provided by considering simultaneously the observed spread of lithium and oxygen (and when 
possible also sodium) abundances for samples of turn-off stars in NGC6752, NGC6121 (M4), and NGC104 (47Tuc), and the helium abundance of their multiple main sequences. 
These observations provide a very powerful test for the accretion scenario, because the observed O, Li and He abundance distributions at the turn off 
can be used to constrain the composition (and mass) of the accreted matter, and the timescales of the polluting stars.
In case of NGC6752 we could not find a physically consistent solution. If early disc accretion does happen, 
observations point towards accretion of gas with a non negligible Li abundance, contrary to the expectations for the ejecta of the 
\lq{natural}\rq\ polluters in this scenario.  
In case of M4, spectroscopic errors are too large compared to the intrinsic spread, to constrain the properties of the accreted matter. 
As for 47Tuc, we could find a physically consistent solution for the abundances of He and O (and Na) in the accreted gas, and predict 
the abundances of these elements in the accreted matter --that will have to be reproduced by 
evolutionary calculations for the polluters and simulations of the global evolution of the intracluster gas-- only if pollution 
happens with timescales of $\sim$1~Myr, hence polluters are objects with masses of the order of several tens of solar masses. 
Accurate spectroscopic measurements of Li and other light elements in dwarf stars in a larger sample of clusters 
are needed to test more comprehensively this scenario.  
}
\authorrunning{Salaris M. \& Cassisi S.}
\titlerunning{Li and O abundances and the early accretion scenario}
  \maketitle

%_____________________________________________________________________

\section{Introduction}

In this last decade, our \lq{classical}\rq\ view of Galactic globular clusters (GGCs) as simple (single-age, single initial chemical composition) 
stellar populations has been severely challenged by a  {\it plethora} 
of photometric and spectroscopic observations \citep[see, e.g.,][and references therein for reviews]{gratton:12, piotto:12}. 
We now know that probably each cluster hosts distinct stellar populations, 
each one with its own chemical abundance pattern characterised
by light element (anti-)correlations (a range of C and O depletions, 
N and Na enhancements, compared the standard $\alpha$-enhanced metal distribution typical of the halo field population) 
and moderate (sometime also large) helium enhancements. 
Depending on the photometric filters \citep[see, i.e.,][]{sbordone:11, cassisi:13}, 
the various sub-populations may be located on separate sequences in the colour-magnitude diagrams, due to the effect of the different abundance 
distributions on isochrones and bolometric corrections.
This broad picture is nowadays denoted as the \lq{multiple population phenomenon}\rq.

The challenge is to understand how distinct sub-populations may have formed in the early stage of the cluster life. 
All the main proposed scenarios 
\citep{decressin, dercole:08,dercole:11,conroy:11, valcarce:11} envisage that the chemical patterns are produced by multiple 
star formation bursts during the early stages of the evolution of clusters. After a first generation (FG) of stars with He abundances and metal 
abundance ratios typical of the halo field population is formed, subsequent  
generations within a cluster (SG stars) originate from 
matter ejected by preexisting FG stars ({\sl polluters}), diluted with material of FG composition not yet involved in star formation episodes.  
The main difference among the various scenarios is related to the nature of the stellar polluters: 
Intermediate-mass asymptotic giant branch stars \citep[see, e.g.,][]{dercole:08}, 
fast rotating massive star \citep[FRMS -- see, e.g.,][]{decressin}, or interacting massive binary star \citep[IMBS -- see, e.g.][]{demink} systems  
are able to eject in the intracluster medium matter with appropriately varied light-element abundances 
and He-enhancement, as a consequence of the nuclear processes associated to high temperature proton captures during H-burning. 
An additional difference related to the nature of the polluters, is the amount of dilution necessary to explain the 
abundances of SG stars; this dilution with pristine material (with FG composition) 
is necessary to reproduce appropriately the observed abundance patterns.

These scenarios encounter some difficulties: {\sl i)} no convincing 
physical mechanism(s) has (have) been yet envisaged that enable a star-forming cluster to keep a significant amount of cold pristine gas; {\sl ii)} 
to explain the actual ratio of SG to FG stars  \citep{carretta:09}, the amount of polluting matter has to be 
extremely large, although only a relatively small fraction of FG stars is contained in the candidate polluters (for a {\sl standard} initial mass 
function). This implies that the present GGCs had to be more massive by a factor 10-100 at birth, at odds with empirical constraints 
based on observations of the GC system in the Fornax dwarf galaxy \citep[see][and references therein]{larsen:12}.
Also, surveys of Galactic and extragalactic massive young clusters with ages between 10~Myr and 1~Gyr --assumed to be the counterpart of GGCs 
at an early evolutionary stage-- do not show any evidence of ongoing star formation \citep{bcd13}, 

A different scenario has been recently envisaged by \citet{bastian:13} --a variant of earlier suggestions by 
\citet{dgc:83} and \citet{thoul:02}. These authors propose that
a fraction of low- and very low-mass FG stars --that we denote as \lq{unenriched}\rq\ stars 
within this scenario-- could accrete, during their fully convective pre Main Sequence (MS) stage, matter ejected 
by IMBS and possibly also FRMSs, and modify their chemical composition to match the observed abundance patterns. 
In this \lq{early disc accretion}\rq\ scenario, the polluting matter is mixed and diluted within the whole star well before the MS phase, 
and no further dilution occurs during 
the following red giant branch phase, in agreement with spectroscopical measurements. In more detail, this scenario envisages the following sequence of 
events: {\sl i)} the cluster becomes gas-free after a single star formation episode with a standard initial mass function, and the more
 massive stars (i.e. both FRMS and IMBS objects) are strongly centrally concentrated as a consequence of mass segregation; {\sl ii)} the matter ejected by 
these fast evolving stars is quickly accreted onto the circumstellar discs surrounding low-mass  objects still on their pre-MS. 
The role of a circumstellar disc is to favour accretion, due to its large cross-section for this process.  

Given that the ejecta of massive stars would be preferentially concentrated in the cluster core, the current number of 
stars with SG composition --that we denote as \lq{enriched stars}\rq\ within this scenario-- 
and the extent of the abundance anticorrelations are modulated by the time spent in the 
cluster core during the pre-MS phase, 
and the efficiency of accretion. According to \citet{bastian:13}, the early disc accretion scenario should be valid for all GGCs, 
excluding only the few ones that show an iron-abundance spread (such as $\omega$ Cen and M22).

In a recent investigation \citep[][hereinafter Paper I]{cs:14} we have tested this scenario 
on NGC2808, that displays a triple main 
sequence with well defined and distinct He abundances. We were able to put tight constraints 
on both the minimum He-mass fraction in the accreted gas, and the accretion efficiency on the \lq{seed}\rq\ fully convective stars. 

In this work we have focused on lithium and oxygen (also Na for one cluster) abundances measured at the turn off (TO) of 
NGC6752, NGC6121 (M4) and NGC104 (47Tuc), three GGCs that also show multiple 
MSs corresponding to well defined He abundances. Lithium is virtually absent 
in the ejecta of FRMS \citep{demink} and IMBS polluters \citep[see, i.e.,][]{decressin_bbbb,fhm10, bmc11}, 
and within the early disc accretion scenario the Li distribution in 
the polluted population is predicted to be determined by the dilution on the preexisting Li with the accreted material  
(and eventually pre-MS depletion due to Li-burning, see next section for details). 
We have investigated whether it is first possible to reproduce simultaneously 
both the abundance of He of the multiple MSs and the measured range of Li abundances at the TO. The combination of these observations provide  
tight constraints on the amount of matter accreted to produce an enriched population with the observed metal distribution and  He   
mass fraction.
For each cluster, given the constraints from the Li abundance analysis, we have then determined 
the oxygen (and sodium for one cluster) mass fraction in the accreted gas, that is required to reproduce the observed spread of 
this element in the same TO stars.
The results provide a crucial additional test of the viability of the early disc accretion scenario.

The next section presents the methodology of our analysis, and is followed by the presentation of the results for the individual clusters.  
A final discussion and conclusions close the paper.

\section{Methods}
\label{simul}

The starting point of our investigation has been the current He abundance distribution of the multiple 
MSs observed in NGC6752, M4 and 47Tuc, as determined by \citet{mpb12}, \citet{mmp13}, and \citet{mmb14}, 
together with the observed spread of ${\rm A(Li)}$ at their TO, measured 
by \citet{shen:10}, \citet{msl11}, and \citet{dlg10}, respectively. We focus on the observed 
Li abundance spread, rather than the absolute abundances, because the spread does not depend 
on the uncertain initial value for the unpolluted stars, possible zero-point offsets of 
the ${\rm T_{eff}}$ scale employed in the spectroscopic analysis (the observed stars span a narrow ${\rm T_{eff}}$ range), and --as we will see below-- 
the uncertain efficiency 
of atomic diffusion, but is a function of the amount of accreted matter in the He-enriched populations.

As shown in Paper~I, 
for each pair of actual MS mass and He abundance Y  (denoted as ${\rm M_{MS}}$ and ${\rm Y_{MS}}$, respectively) 
and an assumed He abundance ${\rm Y_{accr}}$ in the matter accreted by the pre-MS fully convective 
\lq{seed}\rq\ objects --the {\sl free parameter}-- 
it is possible to determine the corresponding amount of mass accreted ${\rm \Delta M_{accr}}$, according to

\begin{equation}
{\rm 
\Delta M_{accr}=M_{MS} \ \frac{Y_{MS}-Y_{ini}}{Y_{accr}-Y_{ini}}
}
\label{eqY}
\end{equation}

where ${\rm Y_{ini}}$ is the Y abundance of the accreting \lq{seed}\rq\ stars.
The condition ${\rm \Delta M_{accr}=M_{MS}}$ --corresponding to the maximum possible accreted mass-- 
provides a firm lower limit for ${\rm Y_{accr}}$, as discussed 
in Paper~I. If ${\rm Y_{MS}}$ along a polluted MS is uniform, 
without any (small) spread, the lower limit for  ${\rm Y_{accr}}$ 
is obviously given by ${\rm Y_{MS}}$; in case of multiple He-rich sequences, this lower limit 
is given by ${\rm Y_{MS}}$ of the most He-rich one. 
  
At the same time, the early accretion scenario predicts that 
the accreting matter is originated by the virtually Li-free ejecta. 
This implies that  
the current surface Li mass fraction of the He-enriched populations --denoted here as ${\rm X(Li)_{MS}}$-- is determined by 
the combined effect of dilution caused 
by accretion of Li-free gas, and the depletion (if any) due to Li-burning in the accreting pre-MS objects.

The typical TO-masses expected for the multiple populations in each of these target clusters are in the 
range ${\rm M_{TO}}\sim$0.75 -- 0.90 ${\rm M_{\odot}}$, for ages of the order of 
11-13~Gyr (the precise value of the clusters' ages does not 
affect any of the quantitative results presented in this section). At the clusters' metallicities 
the predicted pre-MS Li-depletion for stars with these masses is negligible  
\citep{cb97}, and --as we tested with specific calculations-- is independent of the initial He abundance.
This means that the Li pre-MS depletion for a TO star is determined by the depletion within the accreting progenitor 
at the onset of accretion, in the (reasonable, as discussed below) assumption that the duration of the accretion process  
is negligible compared to the typical Li-depletion timescale. 
We denote with $t_f$ the time when  accretion is completed, assumed to be approximately equal to the time 
it started. 

\begin{figure}
\centering
\includegraphics[scale=.4000]{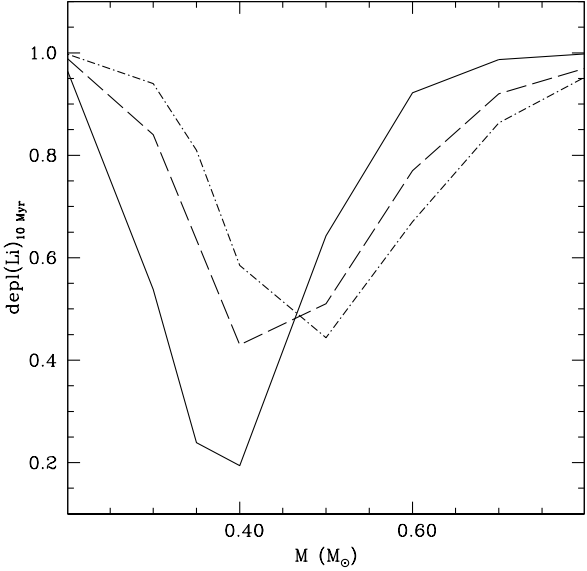}
\caption{Pre-MS Li-depletion factor (final-to-initial abundance ratio) 
at 10~Myr as a function of the stellar mass, from the models by \citet{cb97}, for the metallicities of the  
three selected clusters (NGC6752 -- solid line; M4 -- dashed line; 47Tuc -- dash-dotted line).}
\label{preMSdepl}
\end{figure}

The ratio between the current Li mass fraction for a star on a He-enriched MS, and the initial value of the unpolluted accreting  
population is given by

\begin{equation}
{\rm 
\frac{X(Li)_{MS}}{X(Li)_{ini}}= \frac{depl(Li)_{t_f} \ (M_{MS}-\Delta M_{accr})}{M_{MS}}
}
\label{eqLib}
\end{equation}

where ${\rm (M_{MS}-\Delta M_{accr})}$ is the initial mass of the accreting star, ${\rm X(Li)_{ini}}$ its initial Li mass fraction, 
and ${\rm depl(Li)_{t_f}}$ its Li-depletion factor (the final-to-initial Li abundance ratio) at an age $t_f$.   
By combining Eqs.~\ref{eqY} and ~\ref{eqLib} we obtain, for a given He-rich MS star with mass ${\rm M_{MS}}$ and 
He mass fraction ${\rm Y_{MS}}$, 

\begin{equation}
{\rm 
\frac{X(Li)_{MS}}{X(Li)_{ini}}= depl(Li)_{t_f} \frac{Y_{accr}-Y_{MS}}{Y_{accr}-Y_{ini}}
}
\label{eqLi}
\end{equation}

By employing ${\rm Y_{accr}=Y_{MS}}$ in Eq.~\ref{eqLi}, that is the case of the minimum possible value for ${\rm Y_{accr}}$, 
${\rm (X(Li)_{MS}/X(Li)_{ini})}$=0, an obvious consequence of the fact that the He-rich star is made of just Li-free accreted material. 
For increasingly large values ${\rm Y_{accr} > Y_{MS}}$, the ratio ${\rm (X(Li)_{MS}/X(Li)_{ini})}$ will eventually approach 
${\rm depl(Li)_{t_f}}$.  

We can also generalize Eq.~\ref{eqLi} for the case of an element $i$ not destroyed during the pre-MS and  MS,   
that is present in the accreted matter with a mass fraction ${\rm X(i)_{accr}}$, obtaining

\begin{equation}
{\rm 
\frac{X(i)_{MS}}{X(i)_{ini}}=  \left[\left(\frac{X(i)_{accr}}{X(i)_{ini}}-1\right) \frac{Y_{MS}-Y_{ini}}{Y_{accr}-Y_{ini}}\right] + 1
}
\label{eqZ}
\end{equation}

If the \lq{uncertain}\rq\ efficiency of atomic diffusion in Population~II stars 
\citep[see, i.e., discussions by][and references therein]{korn, msl11} 
does not affect \lq{differentially}\rq\ the surface Li abundances 
of TO stars belonging to the multiple populations of a given clusters, the observed Li abundance spread at the TO can 
essentially be determined 
by comparing the distribution of the ratio ${\rm (X(Li)_{MS}/X(Li)_{ini})}$ in the polluted stars obtained from Eq.~\ref{eqLi},  
with ${\rm (X(Li)_{MS}/X(Li)_{ini})=depl(Li)_{t_f}}$ for the initial unenriched composition.

Our analysis has made use of Monte Carlo simulations for each cluster, whereby we extracted random values for the actual MS 
masses, between 0.1${\rm M_{\odot}}$ and the TO values of both unpolluted and He-enriched MSs,  
inferred from the BaSTI $\alpha$-enhanced ([$\alpha$/Fe]=0.4) isochrones by \citet{basti2}, 
according to a Salpeter initial mass function. This latter choice is irrelevant to our discussion, given that 
to date there are no estimates of the mass functions of the three MSs to compare with, 
and moreover it does not affect the results of our investigation. 
For each of these synthetic MS stars we then determined 
from Eq.~\ref{eqY} the amount of accreted mass required to reproduce the He abundance of the multiple MS,    
for a given value of ${\rm Y_{accr}}$ (the free parameter), and including a small 1$\sigma$=0.003 spread for Y of 
the He-enhanced sequences (we chose this small value to account for the fact 
that the accretion process may not produce precisely the same Y in the stars belonging to the observed multiple MS, and at the same 
time to keep the multiple sequences separated in the colour magnitude diagrams, as observed. 
The exact value of this small spread does not affect any of the results 
presented in the following sections).
The condition ${\rm \Delta M_{accr}=M_{MS}}$ for the most He-rich MS provided the lower limit for ${\rm Y_{accr}}$.

For a given ${\rm Y_{accr}}$ we also assigned to each synthetic star a 
value for the ratio $({\rm X(Li)_{t_f}/X(Li)_{ini}})$, according to Eq.~\ref{eqLi} 
(including a Gaussian 1$\sigma$ spread to mimic the observational errors), and 
determined the value of ${\rm Y_{accr}}$ necessary to match the observed Li abundance spread.
The appropriate  
${\rm depl(Li)_{t_f}}$ was obtained by interpolation amongst the values provided by \citet{cb97},  
for the initial mass and total metallicity [M/H] of the synthetic accreting star.

When the observed Li abundance spread was matched by the synthetic sample with the appropriate choice of ${\rm Y_{accr}}$, we employed 
Eq.~\ref{eqZ} (adding an appropriate observational error) 
to determine the oxygen (and sodium for 47Tuc) abundances (relative to the {\rm unenriched} abundances) 
in the accreted gas, necessary to reproduce the observed O (and Na) abundance spreads. 

Before discussing the results for the individual clusters, we expand upon the issue of atomic diffusion and length of the 
pollution event.

\begin{enumerate}

\item{{\sl Atomic diffusion.} Atomic diffusion during the MS evolution 
can slowly modify the absolute values of the surface Li abundances, especially close to the TO 
(TO stars are hotter,  
and their shallower convective regions induce a faster depletion of the surface Li abundance), 
and potentially also alter the predicted abundance spread. This is because populations with different 
abundances of He will have different TO masses 
(for a fixed age) hence different depths of the convective envelope, that mainly regulates the variation of the surface abundances. 
In case of the largest He abundance differences within each of the three selected clusters (${\rm \delta Y}\sim$0.02-0.03), we have verified 
with calculations of the expected TO masses at an age between 11 and 13~Gyr and the metallicities of our target clusters, that 
the differential effect is at most of the order of $\sim$0.015~dex, for fully efficient atomic diffusion, much smaller than typical spectroscopic errors 
(see next section). Similar result is obtained for other metals like O and Na.
Additionally, spectroscopic 
observations of Li and Fe at the TO and RGB of GGCs show clearly 
that the efficiency of atomic diffusion in these stars \citep[at least the efficiency of diffusion from 
their convective envelopes, see, i.e.][for details]{korn, msl11} is not as expected from the solution of the Burgers equations  
\citep[equations that describe the element transport in a multicomponent fluid associated with diffusion; see, e.g.,][and references therein]{ss}, 
but appears to be moderated by some competing transport mechanism. 
We expect therefore that non-convective element transport mechanisms during the MS 
should not contribute to the Li abundance spread expected at the TO.}

%{\bf Maurizio il paragrafo che segue non lo trovo chiarissimo: In addition to this, M4 and NGC6752 have also Li abundance estimates    
%for samples of lower RGB stars, when the first dredge-up is completed \citep{msl11, msb12}. 
%As discussed for example by \citet{msb12}, the 
%deepening of convection during the dredge-up erases the effect of Li diffusion (if any) along the MS.
%With appropriate evolutionary calculations we derived  
%that the dilution due to the first dredge-up causes in a given cluster  
%differences of Li abundances within $\sim$0.01~dex, 
%for TO masses consistent with ${\rm \delta Y}\sim$0.03 at fixed age, ages of 12-13~Gyr and 
%the metallicity range spanned by the clusters, independently of the efficiency of diffusion. }

%Observationally it is found that for these two clusters the Li abundance spread 
%is essentially the same as at the TO (but obviously not the absolute values).}

\item{{\sl Accretion timescales.} The timing and length of the accretion process from the disk onto 
the \lq{seed}\rq\ object is an important parameter 
to determine the Li abundance in the He-enriched subpopulations. If accretion starts \lq{too late}\rq,  
Li-depletion in the accreting pre-MS fully convective stars 
can be so severe (depending on their mass) that the final abundances are essentially zero. The same is true if 
the accretion process takes a long time and Li is depleted efficiently while the mass slowly increases.

The early disc accretion scenario requires accretion of polluted matter onto the pre-MS objects to be efficient within 
the first $\sim$10~Myr of cluster evolution, when pre-MS low mass stars are still fully convective and 
retain a disk; in our calculations we set this age as an upper limit for $t_f$, that 
corresponds approximately to the lifetime of a 15${\rm M_{\odot}}$ star at the typical metallicities of our clusters 
\citep{gee13}. The lower age limit for the onset of accretion, will be of the order of $\sim 1$~Myr, corresponding to 
the lifetime of a 120${\rm M_{\odot}}$ metal-poor star \citep{gee13}.  

As for the length of the accretion, at the onset of pollution --in the approximation of coeval stars-- 
the disks around the low-mass stars are expected to be depleted of gas, and typical accretion rates at these ages  
for unpolluted stars are observed to be below $\approx 10^{-7}-10^{-8} {\rm M_{\odot} yr^{-1}}$  \citep[see, i.e.,][]{mrd12}. 
However, one expects that the swept intracluster gas replenishes the circumstellar disks and increases 
the accretion rates onto the pre-MS objects  
to values of ${\rm 10^{-4}-10^{-6} M_{\odot} yr^{-1}}$, typical of the earlier main accretion protostellar phase 
\citep{aad14, dsa14}.
With these rates, the accretion time to form the current TO stars would be shorter 
than characteristic Li-depletion timescales, and ${\rm depl(Li)_{t_f}}$ is essentially fixed by the accretor mass and the 
value of $t_f$.
Figure~\ref{preMSdepl} displays the pre-MS Li-depletion factors at 10~Myr 
(the case of $t_f$=1~Myr corresponds to essentially no depletion) used in our calculations for the three target clusters, 
from \citet{cb97} models; notice the strong dependence on both mass and metallicity that must be accounted for, to predict reliable 
${\rm (X(Li)_{t_f}/X(Li)_{ini})}$ ratios. Whenever Li-burning is efficient within a convective region that reaches the surface, 
its photospheric abundance decreases with time. Bearing in mind that 
Li is burned at temperatures  $\sim 2.5 \ 10^6$ K, the observed trend with mass and metallicity for these fully convective objects  
is caused by the interplay of pre-MS evolutionary rate, Li burning timescales, and temperature increase 
within the stellar structures \citep[see][for an analytic treatment of Li depletion in fully convective pre-MS stars]{bil}. 
Uncertainties in the pre-MS Li depletion for the relevant age range arise from variations of the thermal stratification (largely adiabatic) 
of these fully convective models, that is 
mainly affected by uncertainties in the equation of state of the stellar gas, and the surface boundary conditions. 
Regarding the potential effect of rotation in these metal poor stars, 
rotationally-induced MS Li depletion, after the actual TO stars have completed the pre-MS accretion, is 
probably negligible, for rotation rates consistent with the available observations of globular cluster MS stars \citep{pct06}. 
A possible exception is the case of stars born with large rotation rates, that have undergone 
strong magnetic braking at the beginning of the MS \citep{pct06}, and have built up a strong gradient of rotational velocity.
}

\end{enumerate}

\section{Li- and O abundance spread in TO stars}

In this section we present the results of our analysis for the three target clusters.
We have assumed an age t=13~Gyr and [Fe/H]=$-$1.6 for NGC6752, t=13~Gyr and [Fe/H]=$-$1.0 
for M4, t=11~Gyr and [Fe/H]=$-$0.7 for 47Tuc,  
close to the estimates by \citet{sw02} and \citet{cg09} for each cluster. 
The precise values (that fix the TO masses for the various cluster populations and the pre-MS Li 
depletion) are however not critical; for example, a variation of $\pm$1.5~Gyr around the assumed ages 
does not affect at all the results of our analysis.

\subsection{NGC6752}

\begin{figure}
\centering
\includegraphics[scale=.4000]{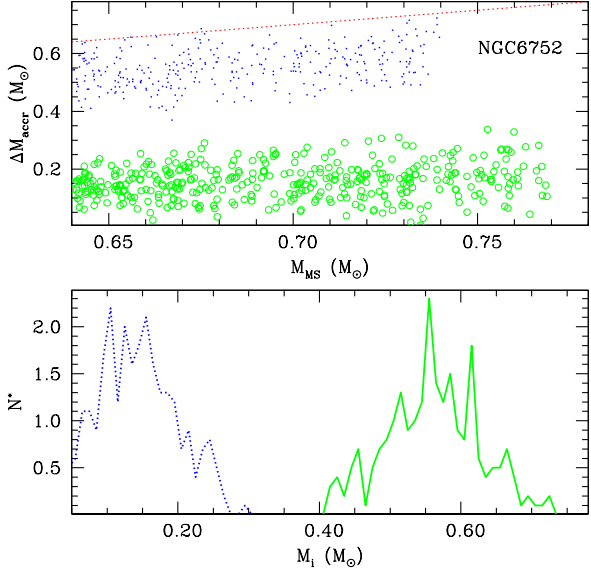}
\caption{{\sl Upper panel:} Mass accreted as a function of the actual MS  
mass (both in solar mass units, in the mass range where separate 
sequences are actually identified observationally) for the two He-rich sequences (dots for the MS with Y=0.275, 
open circles for the MS with Y=0.254) in NGC6752. The dotted
line is the upper bound of the region where the mass of the accretors is larger than zero. 
{\sl Lower panel:} Mass
distribution of the accretors (bin size equal to 0.01 ${\rm M_{\odot}}$). The solid
line corresponds to the seeds of the population with Y=0.254, the dotted line 
is the counterpart for stars with Y=0.275. The quantity ${\rm N^{*}}$ corresponds to the number of stars 
per bin in our simulation, rescaled by a factor 400.
}
\label{N6752_a}
\end{figure}

We considered the triple MS of NGC6752 with ${\rm Y_{MS}}$=0.246 (unenriched), 0.254 and 0.275 (He-enhanced), respectively, 
as derived by \citet{mmp13}. We  
focused on the mass range corresponding to the magnitude interval where the triple MS can be actually separated 
\citep{mmp13}. 
Figure~\ref{N6752_a} displays the results for a simulation with ${\rm Y_{accr}}$=0.280. This 
is the lower limit for the He abundance in the accreted matter to produce the two He-enhanced sequences, 
and has been obtained 
by imposing that the maximum amount of accretion satisfies the relationship ${\rm \Delta M_{accr}=M_{MS}}$ --e.g. 
accretion can be efficient also on brown dwarf objects. 
This lower limit is slightly larger than ${\rm Y_{accr}}$=0.275 --the value expected for ${\rm \Delta M_{accr}=M_{MS}}$ 
in case the He-enhanced sequences do not display any He spread-- because of the small abundance spread assumed for 
the He-enriched sequences. 

As obvious, the two He-enriched populations must have accreted with different efficiencies, 
with the more He-rich sequence displaying larger ${\rm \Delta M_{accr}}$ values 
at fixed ${\rm M_{MS}}$. 
Notice for example 
how the lowest value for the accreting mass ${\rm M_{i}}$ is equal to about 0.4~${\rm M_{\odot}}$ 
for the sequence with ${\rm Y_{MS}}$=0.275. This is because we 
are considering only the portion of the MS where multiple sequences have been detected. 
If the multiple sequence were to coexist down to the H-burning limit, 
the lowest accreting mass would obviously decrease.
 
%If we considered the full stellar mass range down to $\sim$0.1${\rm M_{\odot}}$ --hence assuming the MSs are distinct 
%down to the H-burning limit-- the minimum ${\rm Y_{accr}}$ is 0.290, only slightly different.

%Assuming Mi minimum 0.1Mo
%Observed split MS. minimum Y 0.290
%Assuming split MS down to H-burning limit. minimum Y=0.300

Regarding the observed Li abundances at the TO, spectroscopic measurements 
of 118 stars by \citet{shen:10} derived a substantial 1$\sigma$ spread of 0.15~dex, larger than 
the typical 1$\sigma$=0.09~dex observational error (that we added as a Gaussian error 
to the theoretical ${\rm log(X(Li)_{t_f}/X(Li)_{ini})}$ values). 
We have considered only stars within 0.006${\rm M_{\odot}}$ of the TO masses of the three populations 
(the exact value of this mass range does not change the results), to mimic 
samples of observed TO stars, and 30:40:30 number ratios between objects with  
Y=0.246, 0.254 and 0.275, respectively, in agreement with the estimates by \citet{mmp13}.
We note again that we consider ${\rm (X(Li)_{MS}/X(Li)_{ini})=depl(Li)_{t_f}}$ for the unpolluted population.

\begin{figure}
\centering
\includegraphics[scale=.4000]{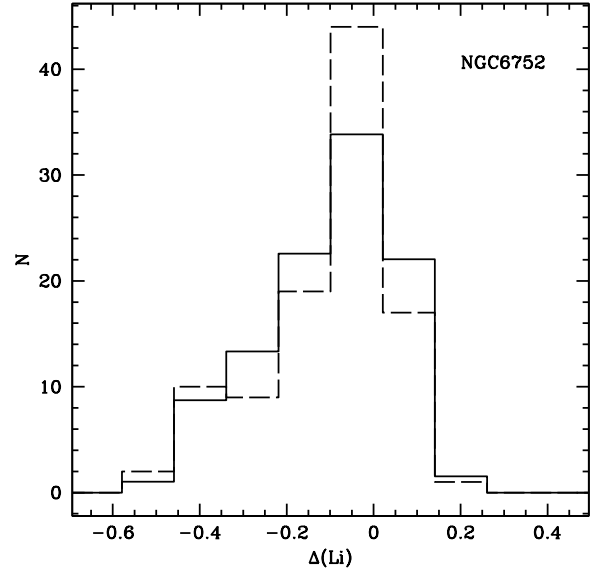}
\caption{Number distribution of theoretical ${\rm \Delta(Li)=log(X(Li)_{MS}/X(Li)_{ini})}$ 
values at the TO for $t_f$=10~Myr, compared to observations of Li abundances in NGC6752.   
The observed abundances have been shifted to make their mean value match the mean value of 
${\rm \Delta(Li)}$. The theoretical number distribution has been rescaled to match the total number of stars with 
Li measurements. The histogram bin size is equal to 0.12~dex.
}
\label{N6752_b}
\end{figure}

With $t_f $ equal to either 1~Myr or $t_f $=10~Myr, 
He abundances ${\rm Y_{accr}}$ equal to 0.31 or 0.35 are required for ${\rm (X(Li)_{t_f}/X(Li)_{ini})}$ 
values of the whole TO sample (unenriched and enriched sub-populations) to reproduce the observed  
1$\sigma$ spread. 
Figure~\ref{N6752_b} compares the number distribution of ${\rm log(X(Li)_{t_f}/X(Li)_{ini})}$ values for the 
best match sample of synthetic TO stars (200 objects)  
including all three populations (30:40:30 ratios) with the empirical measurements, for the 
case of ${\rm Y_{accr}}$=0.35 and $t_f$=10~Myr; notice that the observed abundances 
were shifted to make their mean value match the mean value of ${\rm log(X(Li)_{t_f}/X(Li)_{ini})}$. 
The theoretical 
number distribution has been rescaled to match the observed total number of stars with spectroscopic measurements of Li.

Additionally we tested with a KS-test that the whole Li distribution of the best match synthetic samples  
and the observed one (absolute values shifted as described 
before) are consistent.  
As customary, we considered the distributions to be different when the KS-test 
returns a 95\% probability of a significat difference. Here, and for 
all cases discussed below, this probability is always around 50\% or less.

The increase of ${\rm Y_{accr}}$ for $t_f $=10~Myr is the logical consequence of pre-MS 
Li-depletion affecting low mass stars at this age, that demands a smaller dilution due to 
accretion (hence less accreted mass and higher ${\rm Y_{accr}}$) to reproduce the observed Li spread. 
The mean mass ratios between actual and initial mass of the TO stars are equal to 1.13 and 1.08 for the Y=0.254 population, 
and 1.85 and 1.38 for the Y=0.275 population.
The mean values of ${\rm log(X(Li)_{t_f}/X(Li)_{ini})}$ for the whole TO sample are equal to $-$0.12 and $-$0.10 for the two choices of $t_f$. 

The spectroscopic measurements by \citet{shen:10} provided also O abundances for TO stars, with a typical 
1$\sigma$=0.14~dex error on the individual estimates. 
The resulting 1$\sigma$ spread of the O abundance distribution is equal to 0.28~dex. 
We have attempted to match 
the observed O abundance spread with the range of ${\rm log(X(O)_{MS}/X(O)_{ini})}$ 
obtained from Eq.~\ref{eqZ} (considering ${\rm log(X(O)_{MS}/X(O)_{ini})=0}$ for the unpolluted stars) 
by calibrating the ratio ${\rm (X(O)_{accr}/X(O)_{ini})}$ 
for ${\rm Y_{accr}}$ fixed to the values obtained from the Li abundance spread analysis. 
We also included a Gaussian 1$\sigma$=0.14~dex random errors for all cluster populations 
to mimick the observations.

The maximum spread obtained from our simulations corresponds to the case of no oxygen in the accreted matter, 
that causes the maximum possible dilution in the He-enhanced populations. Even in this extreme case we 
could not match the observed  1$\sigma$=0.28~dex spread for both choices of $t_f$. We obtained instead 
values equal to 0.18~dex for $t_f$=1~Myr and and 0.15~dex for $t_f$=10~Myr.

In the comparison of the observed Li and O abundances with our simulations 
we have employed \citet{shen:10} results, discarding objects with only upper limit abundance estimates.
As an additional check, we have also used the much smaller sample of Li measurements for 9 TO stars 
by \citet{pbm05}. The abundance spread is the same (and the 1$\sigma$ 
observational error is also very similar) as the case of \citet{shen:10}. These authors discuss also 
O abundances for the same stars taken from \citet{cgl05}. 
Using this smaller set of Li and O abundances we were again unable to match consistently 
both observed Li and O abundance ranges with our simulations.

\subsection{M4}

Deep optical and near-infrared photometries of M4 have disclosed a double sequence at the bottom of the MS  
\citep{mmb14}, corresponding to initial Y values equal to 0.25 (unenriched) and 0.27 (He-enhanced). 
The split MS cannot be traced up to the TO region, but we make the reasonable assumption that the whole MS is 
separated into these two components.

\begin{figure}
\centering
\includegraphics[scale=.4000]{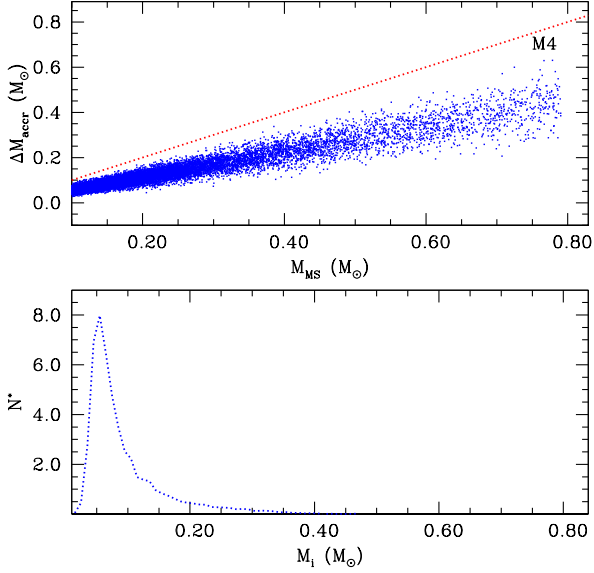}
\caption{As Fig.~\ref{N6752_a} but for the He-rich MS of M4. The quantity ${\rm N^{*}}$ corresponds to the number of stars 
per bin in our simulation, rescaled by a factor 400.}
\label{M4_a}
\end{figure}

Monte Carlo simulations using Eq.~\ref{eqY} to determine 
${\rm \Delta M_{accr}}$ as a function 
${\rm Y_{accr}}$, provide a lower limit ${\rm Y_{accr}}$=0.285 (see Fig.~\ref{M4_a}) to 
produce the observed He-enhanced sequence.

\begin{figure}
\centering
\includegraphics[scale=.4000]{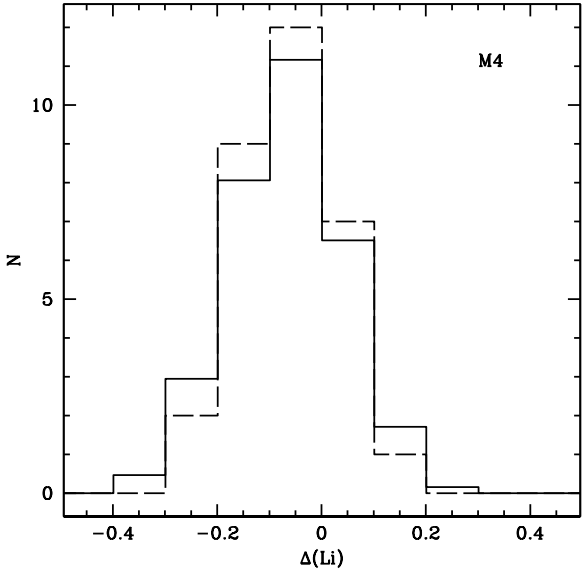}
\caption{As Fig.~\ref{N6752_b} but for the TO region of M4. The histogram bin size is equal to 0.10~dex.}
\label{M4_b}
\end{figure}

Spectroscopic observations by \citet{msl11} of a sample of 31 TO stars determined a 1$\sigma$=0.10~dex 
Li abundance spread, consistent with the size of the estimated observational errors.  
Our simulations that make use of Eq.~\ref{eqLi} (including a 
Gaussian 1$\sigma$=0.10~dex spread, to mimic the observational errors) could reproduce the observed Li spread 
with ${\rm Y_{accr}}$ equal to 0.39 and 0.45, 
for $t_f $=1~Myr and 10~Myr, respectively (see Fig.~\ref{M4_b} for a sample of 200 objects and the case $t_f$=10~Myr). 
A KS-test also confirmed the agreement between synthetic and observed Li abundance samples for these parameters 
(with the measured abundances shifted to match the mean value of ${\rm log(X(Li)_{t_f}/X(Li)_{ini})}$).  
This result does not depend critically on the number ratio between the two populations \citep[that we assumed to be 40:60 between unpolluted 
and He-enhanced stars, following][]{mmb14} 
given that for the spread to be equal to the observational errors, the 
Li abundance distribution --in our case the distribution of ${\rm (X(Li)_{t_f}/X(Li)_{ini})}$ ratios-- 
must be essentially the same for both groups of stars \citep[see also][]{msl11}. 
The relatively high  ${\rm Y_{accr}}$ is required to keep 
${\rm \Delta M_{accr}}$ small, and causes a very small Li dilution within the He-enriched population, to match the observations.
As a consequence, the mean mass ratio between actual and initial mass of the He-enhanced TO stars is close to unity, 
equal to 1.16 and 1.11, for $t_f$=1~Myr and 10~Myr respectively. The corresponding 
mean values of ${\rm log(X(Li)_{t_f}/X(Li)_{ini})}$ are $-$0.04 and $-$0.05~dex.   

The analysis by \cite{msl11} provided also O abundances for the TO stars, characterized by a 
1$\sigma$=0.15~dex error on the individual estimates; the resulting 
O abundance distribution displays a 1$\sigma$ spread also equal to 0.15~dex.
Calculations as those done for NGC6752 revealed that the data cannot provide any constraint on 
the O abundance in the polluting matter, given that  
the observed spread is dominated by the spectroscopic error. Any value of the oxygen abundance in the accreted matter 
between the unpolluted value ${\rm X(Li)_{ini}}$ and zero (the accreting matter must be O-depleted to reproduce the observed 
abundance patterns of enriched stars) reproduces the observed spread for both choices of $t_f$, 
once the observational error is included in the theoretical simulations, and satisfies simultaneously the KS-test.

\subsection{47Tuc}

\begin{figure}
\centering
\includegraphics[scale=.4000]{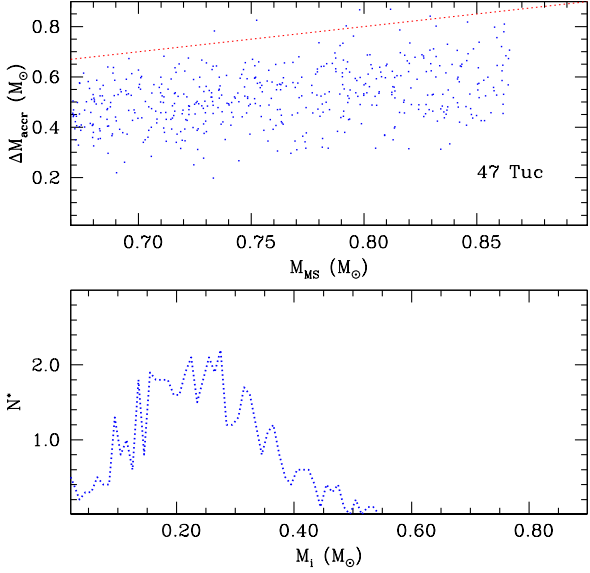}
\caption{
As Fig.~\ref{N6752_a} but for the He-rich MS of 47Tuc. The quantity ${\rm N^{*}}$ corresponds to the number of stars 
per bin in our simulation, rescaled by a factor 10.}
\label{47Tuc_a}
\end{figure}

The study by \cite{mpb12} has disclosed a double MS for 47Tuc, 
corresponding to a standard initial He abundance Y=0.25, and a He-enhanced one with Y=0.265.
We have determined a lower limit ${\rm Y_{accr}}$=0.270 (when considering the small abundance spread along the He-enhanced sequence)  
to produce the He-enhanced MS (see Fig.~\ref{47Tuc_a}, that displays the mass interval where the double MS 
is detected).

\begin{figure}
\centering
\includegraphics[scale=.4000]{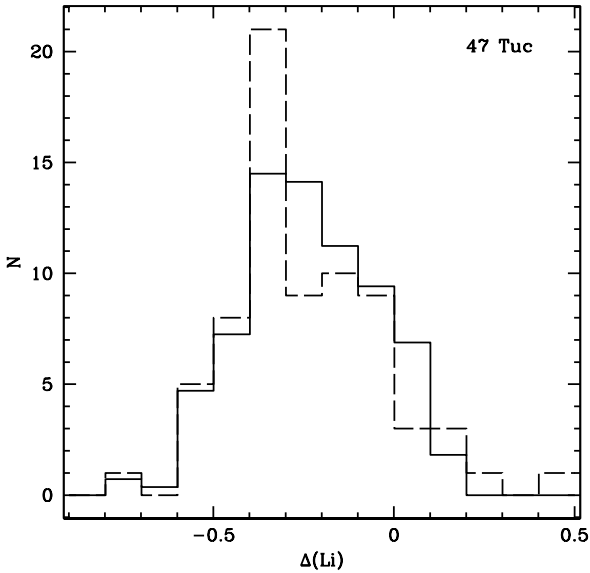}
\caption{As Fig.~\ref{N6752_b} but for the TO region of 47Tuc. The histogram bin size is equal to 0.10~dex.}
\label{47Tuc_b}
\end{figure}

Regarding Li abundances, the spectroscopic analysis by \citet{dlg10}  
determined a spread 1$\sigma$=0.20~dex for a sample of 71 TO stars, much larger than the average observational error 
(1$\sigma$=0.08~dex)\footnote{The very recent study by \citet{dkb13} of Li abundances at the TO of this cluster  
retrieves the same abundance spread, with a typical 1$\sigma$ spectroscopic error also 
equal to \citet{dlg10}. The absolute values --irrelevant to our analysis-- are however smaller by about 0.3~dex on average, 
due to the use of a 3D NLTE spectral synthesis.}.   
Our simulations (that include a 1$\sigma$=0.08~dex random Gaussian observational spread) 
were able to reproduce the observed Li spread with ${\rm Y_{accr}}$, 
equal to 0.278 and 0.30, for $t_f$=1~Myr and 10~Myr, respectively   
(see Fig.~\ref{47Tuc_b} for a sample of 200 synthetic objects and the case $t_f$=10~Myr). We considered a 30:70 
ratio between He-normal and He-enhanced TO stars, according to the observed ratio of 
O-normal to O-depleted stars in the TO sample with Li measurements by \citet{dlg10}. 
A KS-test confirms the agreement between the Li abundance synthetic samples and the observed one 
(with the measured abundances suitably shifted as described for NGC6752 and M4).
The mean value of ${\rm log(X(Li)_{t_f}/X(Li)_{ini})}$ is equal to $-$0.18 for both assumptions 
about $t_f$, and the corresponding mean mass ratios between actual and initial mass of the He-enhanced TO stars 
are equal to 2.23 and 1.43, respectively.

\begin{figure}
\centering
\includegraphics[scale=.4000]{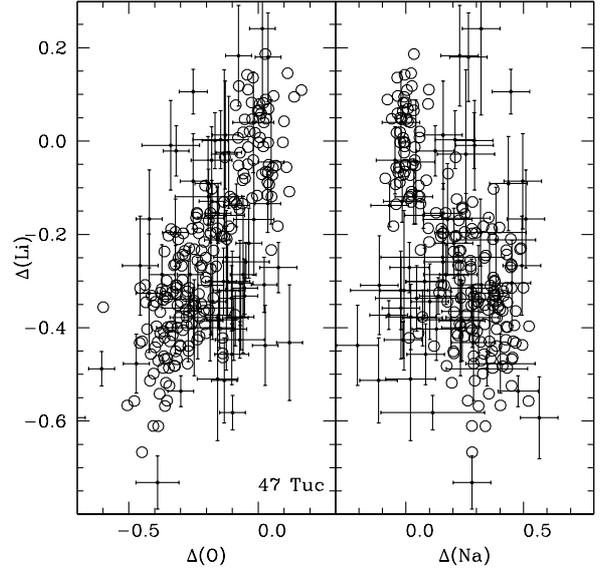}
\caption{
${\rm \Delta(Li)=log(X(Li)_{MS}/X(Li)_{ini})}$ vs the corresponding values for oxygen (left panel) and sodium (right panel), 
for the 47Tuc best match synthetic sample with ${\rm Y_{accr}}$=0.278, ${\rm (X(O)_{accr}/X(O)_{ini})}$=0.1 and ${\rm (X(Na)_{accr}/X(Na)_{ini})}$=3.
The observed Li, O and Na abundances are also displayed (symbols with error bars), shifted to match the mean values 
of  ${\rm \Delta(Li)}$, ${\rm \Delta(O)}$, and ${\rm \Delta(Na)}$.}
\label{47Tuc_anti}
\end{figure}

\citet{dlg10} study provided also abundances of O and Na for TO stars, 
characterized by a 1$\sigma$ spread equal to 0.16~dex and 0.18~dex, respectively. 
Using Eq.~\ref{eqZ} together with random Gaussian 1$\sigma$ errors equal to 0.06~dex for O and 0.10~dex 
for Na (equal to the spectroscopic errors), we found that for $t_f$=1~Myr (${\rm Y_{accr}}$=0.278), both 
O and Na abundance spreads could be reproduced for chemical abundances in the ejecta equal to 
${\rm (X(O)_{accr}/X(O)_{ini})}$=0.1 and ${\rm (X(Na)_{accr}/X(Na)_{ini})}$=3.
We display in Fig.~\ref{47Tuc_anti} the observed Li-O and Li-Na abundance-abundance patterns, compared 
to the results from this synthetic sample with ${\rm Y_{accr}}$=0.278. The simulation predicts 
log${\rm (X(i)_{MS}/X(i)_{ini})}$ ratios of Li, O and Na for each synthetic star, and the observed abundances have 
been shifted to match the mean values of these ratios. A KS-test also confirms the agreement between the 
abundance of both Na and O of the synthetic samples, and the observed values (shifted as in  Fig.~\ref{47Tuc_anti}).

In case of  $t_f$=10~Myr (${\rm Y_{accr}}$=0.30) it was not possible to match the 
observed spread of O, even assuming ${\rm (X(O)_{accr}/X(O)_{ini})}$=0 (we obtained a maximum 
1$\sigma$ spread equal to 0.10~dex), whilst the Na abundance spread was 
matched with ${\rm (X(Na)_{accr}/X(Na)_{ini})}$=5.  
 
\section{Discussion and conclusions}

An attractive feature (by no means the only one) of the early disc accretion 
scenario proposed by \citet{bastian:13} to explain the GGC multipopulation phenomenon, is that it can be easily 
subject to stringent tests when abundance estimates are available for He, Li and elements involved in the observed 
anticorrelation patterns.
We have expanded upon our first analysis of Paper~I, by 
considering three clusters (NGC6752, M4 and 47Tuc) with abundance estimates for Li and O (also Na for 47Tuc) in samples of TO stars, and with 
multiple MSs with well defined He abundances.

The He abundances along the MS of each clusters, together with the measured range of Li, have allowed us to put tight constraints on the 
amount of matter accreted by enriched stars, and its He mass fraction.
These latter abundances are well within the values expected in the ejecta of IMBS or FMRS polluters.
For the amount of accreted matter fixed by the Li analysis, 
we have then determined its O (and Na for 47Tuc) content, using the observed spread in TO stars as constraint. 

We could not draw any conclusion from M4 O abundance measurements, because the observed spread is dominated by the observational 
error. Any abundance in the accreted gas between the unenriched value and zero can satisfy the observations.

As for 47Tuc, we found that the observed abundance spread of O and Na can be simultaneously matched when $t_f$=1~Myr 
and ${\rm Y_{accr}}$=0.278 --very close to the lower limit for ${\rm Y_{accr}}$ discussed in Sect. 3.3-- if ${\rm (X(O)_{accr}/X(O)_{ini})}$=0.1 and 
${\rm (X(Na)_{accr}/X(Na)_{ini})}$=3. This match cannot be achieved for 
$t_f$=10~Myr. This points towards stars with mass equal to several tens of solar masses as the source of the pollution. 
It has to be seen whether detailed IMBS and FRMS calculations  
for the appropriate cluster metallicity provide these chemical abundances (and enough matter with this composition) in the ejecta 
of models with the right evolutionary timescales, and whether simulations of the 
global evolution of the intracluster gas predict the right amount of accretion 
with this chemical composition.

The situation is much more unfavourable for NGC6752. The oxygen spread cannot be matched for any choice of $t_f$, 
even for oxygen-free accreted gas.
This result confirms the conclusions by \citet{shen:10}, whose data we have employed in our analysis.
The problem is that the measured Li abundance spread is (even considering differences in the spectroscopic errors) 
smaller than the O abundance range. Given that the accreted matter has to be Li-free (FMRS and IMBS are the polluters in this scenario) 
the maximum dilution of oxygen in the final enriched composition cannot exceed the dilution of Li. It is therefore impossible to satisfy 
simultaneously both observed  O- and Li abundance spread. Essentially, to reproduce the O abundance range, a larger amount 
of accreted matter (hence lower ${\rm Y_{accr}}$) is needed.
An increased efficiency of pre-MS Li-burning in the \lq{seed}\rq\ objects --due for example to short bursts of very high accretion rates,  
as explored at solar metallicity by \citet{bc10}-- would exacerbate the problem, because one would need a smaller dilution --hence less 
matter accretion-- to reproduce the Li abundance range.

In general, both O and Li abundance measurements are dominated by errors in ${\rm T_{eff}}$ and in equivalent width measurement 
\citep{shen:10}. 
It is important to stress that the {\sl relative} abundances of Li and O are relevant to this analysis, not the absolute values, and also 
that we are dealing with objects (TO stars) covering a narrow range of ${\rm T_{eff}}$ and surface gravity. 
As discussed by \citet{shen:10}, there are for example uncertainties in the absolute values of the oxygen abundances obtained 
from the OI (777.1-777.5 nm) triplet lines used for the abundance determination, 
but the relative values (hence the measured spread) are expected to be much more reliable.

Notice also that the same conclusions regarding NGC6752 are found when using the smaller sample of TO stars studied by 
\citet{pbm05}. These same authors also noticed that in their stars the observed spread of O abundances is larger than the Li spread.

A possible solution of this problem envisages pre-mixing of IMBS and/or FRMS ejecta  
with non Li-free gas. We performed some test simulations 
and derived that agreement with the observed Li and O spread, and the He abundances along the triple MS, can be achieved  
with ${\rm Y_{accr}}\sim$0.285 --close to the minimum possible value-- $t_f$=1~Myr 
(hence no pre-MS Li-depletion), ${\rm (X(O)_{accr}/X(O)_{ini})}\sim$0.01 and ${\rm (X(Li)_{accr}/X(Li)_{ini})}\sim$0.3. 
If the gas involved in the pre-mixing before accretion is characterized by an abundance of Li 
equal to the unenriched composition, about 50\% of the matter polluting pre-MS stars had to come from this external source.
Notice also the extremely low oxygen abundance to be accreted. Of course, the need of an   
(additional) rather {\sl ad hoc} dilution process introduces an unwelcomed fine-tuning in the early disc accretion scenario  
to explain NGC6752.

To summarize, we have shown how TO abundances of Li and O and estimates of He along the MS can provide stringent tests 
of the early disc accretion scenario. In particular, it is possible to put tight constraints on the amount of matter 
accreted by the unenriched \lq{seed}\rq\ stars, He and O content of the accreted gas, the timescales of the polluters.

Of the three clusters investigated, the abundances in NGC6752 appear to be difficult to reproduce.
Accurate TO spectroscopic measurements, and estimates of the He abundance along the MS of more clusters 
are needed to test more comprehensively this scenario.  

%We close this section by observing that measurements of the Li abundance spread below the TO can provide further tests/constraints for this scenario,  
%in case accretion happens at ages of the order of $t_f$=10~Myr. 
%Just as an example, 
%in case of NGC6752, one magnitude below the TO in the F336 filter (actual masses lower by about 0.06${\rm M_{\odot}}$ compared to the TO values) 
%the simulations discussed in the previous section predict a 1$\sigma$ Li spread $\sim$0.20~dex for ${\rm Y_{accr}}=0.35$ ($t_f$=10~Myr), larger than
%1$\sigma$=0.15~dex observed at the TO.
%The difference between the results for the two different magnitude levels along the MS 
%is due to the variation of the accreting initial mass and the associated Li-depletion before the accretion started --no variation is 
%expected for $t_f$=1~Myr, or for the alternative scenario of multiple stellar generations.
%The extent of this spread is, again, not affected by atomic diffusion (if fully efficient), as we have verified with appropriate evolutionary
%models of the final post-accretion objects.

\begin{acknowledgements}
We warmly thank P. Bonifacio for providing us with the results for the lithium and oxygen abundances in NGC6752, 
A. Mucciarelli for the oxygen abundances in M4, S. Longmore for enlightening discussions about 
star formation and disc accretion, and an anonymous referee for several comments that helped to 
improve the presentation of our results. 
SC acknowledges financial support
from PRIN-INAF 2011 "Multiple Populations in Globular Clusters: their
role in the Galaxy assembly" (PI: E. Carretta), and from PRIN MIUR 2010-2011,
project \lq{The Chemical and Dynamical Evolution of the Milky Way and Local Group Galaxies}\rq, prot. 2010LY5N2T (PI: F. Matteucci). 
\end{acknowledgements}

\bibliographystyle{aa}
%\bibliography{salarisLi}

\end{document}